\documentclass[11pt]{article}

\usepackage{amsfonts}
\usepackage[centertags]{amsmath}
\usepackage{amssymb}
\usepackage{tikz}
\usetikzlibrary{matrix}
\usetikzlibrary{decorations.markings,calc,shapes}
\usetikzlibrary{positioning}
\usepackage{cite}
\usepackage{psfrag}
\usepackage{graphicx}
\usepackage{bbm,bm}
\usepackage{epsfig, multicol}
\usepackage{appendix}
\allowdisplaybreaks
\begin{document}

\topmargin 0pt
\oddsidemargin 0mm
\def\be{\begin{equation}}
\def\ee{\end{equation}}
\def\bea{\begin{eqnarray}}
\def\eea{\end{eqnarray}}
\def\ba{\begin{array}}
\def\ea{\end{array}}
\def\ben{\begin{enumerate}}
\def\een{\end{enumerate}}
\def\nab{\bigtriangledown}
\def\tpi{\tilde\Phi}
\def\nnu{\nonumber}
\newcommand{\eqn}[1]{(\ref{#1})}

\newcommand{\half}{{\frac{1}{2}}}
\newcommand{\vs}[1]{\vspace{#1 mm}}
\newcommand{\dsl}{\pa \kern-0.5em /} 
\def\a{\alpha}
\def\b{\beta}
\def\g{\gamma}\def\G{\Gamma}
\def\d{\delta}\def\D{\Delta}
\def\ep{\epsilon}
\def\et{\eta}
\def\z{\zeta}
\def\t{\theta}\def\T{\Theta}
\def\l{\lambda}\def\L{\Lambda}
\def\m{\mu}
\def\f{\phi}\def\F{\Phi}
\def\n{\nu}
\def\p{\psi}\def\P{\Psi}
\def\r{\rho}
\def\s{\sigma}\def\S{\Sigma}
\def\ta{\tau}
\def\x{\chi}
\def\o{\omega}\def\O{\Omega}
\def\k{\kappa}
\def\pa {\partial}
\def\ov{\over}
\def\nn{\nonumber\\}
\def\ud{\underline}
\def\qq{$Q{\bar Q}$}
\begin{flushright}
%
\end{flushright}
\begin{center}
{\Large{\bf Multipartite Purification, Multiboundary Wormholes and Islands in AdS$_{3}$/CFT$_{2}$ }}

\vs{10}

{Aranya Bhattacharya\footnote{E-mail: aranya.bhattacharya@saha.ac.in}}

\vs{4}

{\it Saha Institute of Nuclear Physics\\
1/AF Bidhannagar, Calcutta 700064, India}

\vs{4}

{\rm and}

\vs{4}

{\it Homi Bhabha National Institute\\
Training School Complex, Anushakti Nagar, Mumbai 400085, India}
\end{center}

\begin{abstract}
    The holographic duals of Entanglement of Purification through the Entanglement Wedge Cross Section has been a well-discussed topic in the literature recently. More general entanglement measures involving multipartite information and their holographic duals have also been proposed. On the other hand, the recent intriguing program reproducing the Page Curve in Black hole entropy using the notion of islands has also been an obvious issue of attraction. A toy model involving Multiboundary wormholes in AdS$_{3}$ was able to capture many interesting facts about such calculations. In such a toy model, the notion of islands was intuitively connected to quantum error correction. We try to bridge the ideas of the two programs especially in AdS$_{3}$/CFT$_{2}$ and give a description of the islands in terms of multipartite entanglement of purification. This clarifies a few simplified assumptions made while describing the toy model and also enables us to understand the familiar information paradox within the framework of the same model.
\end{abstract}
\newpage
\tableofcontents
\section{Introduction:}
In the last several decades, the black hole information \cite{PhysRevD.13.191} problem has resurfaced more often than not in the studies of BH thermodynamics as well as in the studies of black holes through holography. Holography has proved to be a very insightful candidate in many such studies. The way to tackle the information problem in holography has been boosted by quantum information-theoretic studies \cite{Bombelli:1986rw, Srednicki:1993im, Holzhey:1994we,Rangamani:2016dms, Calabrese:2004eu, Eisert:2008ur, 
Nishioka:2009un, Takayanagi:2012kg} done in holography, where the stand out contribution has come in from the seminal Ryu-Takayanagi \cite{Ryu:2006bv, Jafferis:2015del, Lewkowycz:2013nqa} conjecture relating bulk codimension-two minimal surfaces to boundary entanglement entropy.

Recently, in \cite{Takayanagi:2017knl}, the authors have conjectured bulk counterparts of more general information-theoretic quantities like the entanglement of purification. This has been studied both in gravitational setups as well as in free field theories \cite{Caputa:2018xuf, Bhattacharyya:2019tsi, Bhattacharyya:2018sbw, Akers:2019gcv}. In the gravitational ones, the holographic EoP is conjectured to be related to the entanglement wedge cross-section. Such results include the study of pure AdS, BTZ black holes as well as time-dependent scenarios. A study of few other quantum information-theoretic quantities e.g; multipartite entanglement of purification \cite{Bao:2018gck, Umemoto:2018jpc, Bao:2017nhh, Kusuki:2019rbk, Umemoto:2019jlz}, reflected entropy \cite{Dutta:2019gen, Bao:2019zqc, Chu:2019etd, Moosa:2020vcs, Jeong:2019xdr, Kusuki:2019evw} have been motivated by the EoP computations.

On the other hand, since last one year or so, the information paradox problem has also gone through an intriguing turn of events due to several interesting papers \cite{Almheiri:2019yqk, Almheiri:2019qdq, Almheiri:2019hni, Almheiri:2019psf, Akers:2019nfi}, where the authors try to describe the famous Page curve using the quantum, corrected holographic entanglement entropy for evaporating black hole geometries (note that these are time-dependent scenarios). The motivation for these works is also somewhat related to the entanglement wedge \cite{Harlow:2018fse} studies done both in holographic as well as in nonholographic setup (using quantum maximin surfaces)\cite{Akers:2019lzs}. The results have attracted a lot of interest in the community since it has been able to explain the long-lasting problem of describing the Page curve convincingly. 

Surprisingly, the two above-mentioned studies have not yet been discussed together. To be precise, that is the motivation and distant goal of this paper, i.e; to point out the similarities in the study of EW cross-section (conjectured EoP) and the Page curve study for black holes.

In this paper, we discuss the possible connections between the two studies in the case of AdS$_{3}$. We use the concepts of the holographic dual of multipartite entanglement of purification for states in the boundary of pure AdS$_{3}$. We use these concepts in the toy model of evaporating black hole \cite{Akers:2019nfi} where the evaporating black hole is initially treated as a big black hole whereas the emitted Hawking Quantas are realized by smaller black holes in AdS$_{3}$. All of the black holes are connected by a multiboundary wormhole. An important fact in this connection would be the realization of multiboundary wormhole(MBW) in AdS$_{3}$ as quotients of pure AdS$_{3}$, where the boundaries are identified by removing semicircles and introducing orientation reversing isometries in a timeslice of AdS$_{3}$. 

The CFT at each of the boundaries would be subregions of the CFT that we deal with while discussing multipartite entanglement of purification and there, the horizon lengths are independent of each other. Therefore, we can tune the sizes of the black holes (Hawking Quantas would be smaller black holes compared to the one that is emitting them) by choosing the parameters(/intervals in the boundary CFT of pure AdS$_{3}$) and fixing them in such a way that the smaller black holes would be of comparable sizes. The bigger black hole keeps decreasing in size while creating more and more such smaller black holes which increases the exits of the wormhole.  In the picture described in the toy model, the preferred HRT surface change over time with the inclusion of a shared interior. This shared interior is conjectured to be the analog of the nontrivial islands mentioned in Maldacena's papers.

We would try to get an understanding of the islands from the perspective of multipartite EoP. We would make comments upon the shared interior from the point of view of quantum error connection \cite{Harlow:2018fse, Pastawski:2015qua, Dong:2016eik}, but by taking a detour through entanglement of purification. This is not very surprising since both of the programs are heavily dependent on the ideas of entanglement wedge reconstruction and nesting. Our comments relate the body of a muliboundary wormhole to a geometric pure state construction. We further figure out how in such a model, there are two sides to the whole story. The classical picture gives us an intuitive understanding of the islands and helps to reproduce the Page curve. But the quantum version of the extremal surface again gives back the familiar paradox addressed by Hawking. Finally we give resolutions through which one can understand how to deal with the problem both in the toy model as well as in the entanglement of purification case.

The rest of the paper is constructed in the following way. We give a brief review of the purification program in section \ref{section2}. In section \ref{section3}, we review the basic ideas that have been instrumental in the derivation of the Page curve for the Black holes and the Hawking radiation. We also discuss the toy model which motivated this work. We discuss the multiboundary wormhole constructions in AdS$_{3}$ by quotienting AdS$_{3}$ with several pictures that describe the situation in a simple way. We would find that it is actually necessary to review the two topics in such a manner as it helps us to get hold of the connections made in the later half. In section \ref{section4}, we compare the two programs to point out the similarities and differences and how do the connections can feed each other in several interesting ways. Finally, in section \ref{section5}, we discuss resolutions to strengthen the connection, a few open problems and future directions that one can pursue.


\section{A Note on Entanglement of Purification:}{\label{section2}}

Purification is the process of making a mixed state pure. There are numerous measures in quantum information theory, most of which are sensitive to the state in hand being a pure one. A pure state is a state for which one can use the standard ket ($|\psi\rangle$) notation of quantum mechanics. The density matrix of a pure state ($\rho_{pure}=|\psi\rangle \langle \psi |$) follows a simple relation,

\begin{equation}
    Tr[\rho_{pure}]=Tr[\rho_{pure}^{2}] = 1.
\end{equation}
 A mixed state has many descriptions from the perspective of quantum mechanics, the most famous one being the following,
 a mixed state is a classical probabilistic mixture of all the possible outcome states. It only has a representation in terms of the density matrix ($\rho_{mixed} = \sum_{i} p_{i} |\psi_{i}\rangle \langle \psi_{i} |$) and does not have a simple ket description.
 
 \begin{equation}
      Tr[\rho_{mixed}]= 1, Tr[\rho_{mixed}^{2}] < 1.
 \end{equation}

The standard way to purify a mixed state is to add auxiliary system with the mixed state where the total state after adding the auxiliary system becomes a pure state and the mixed state becomes a particular reduced state after tracing out a few degrees of freedom from the purified state. But , for a single mixed state, there might exist more than one way of purification. There, a particular one among them is chosen with respect to the information theoretic measure one wants to calculate in a given scenario.

\subsection{Definitions and Properties:}\label{PuriQM}
Entanglement of purification, as the very name suggests, is related to purification of a mixed quantum state. The precise definition of entanglement of purification between A and B for a bipartite mixed state AB ($= A \cup B $) is the minimal entanglement entropy between $A A^{\prime}$ and $B B^{\prime}$, where $A^{\prime}$ and $B^{\prime}$ are auxiliary systems added to make the whole state $A A^{\prime} B B^{\prime}$ pure. 

\begin{figure}[t]
\centering
\includegraphics[width=0.49\textwidth]{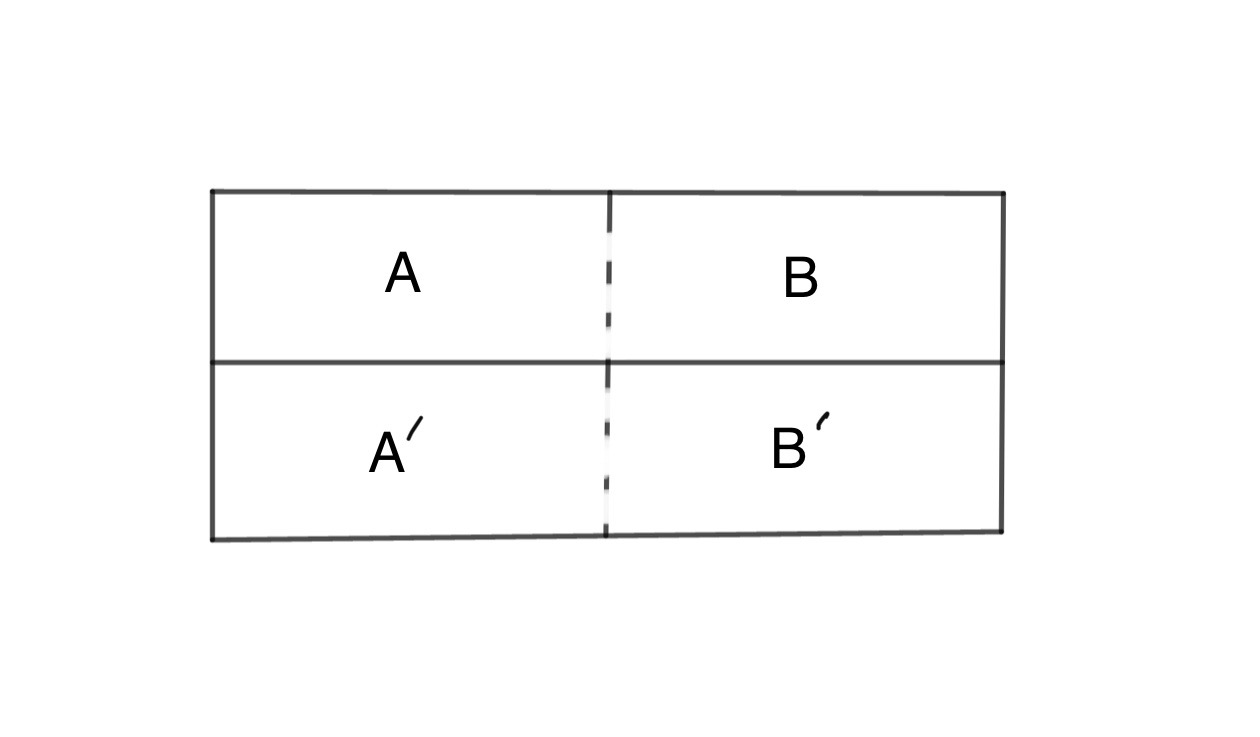}
\caption{A schematic diagram of bipartite purification ($A A^{\prime} B B^{\prime}$ forms a pure state and EoP is entanglement between $A A^{\prime}$ and $B B^{\prime}$)}
\label{fig:1}
\end{figure}

Similarly , multipartite entanglement of purification is where instead of a bipartite mixed state AB, we start with a multipartite mixed state $A_{1} A_{2} ..... A_{n}$ and add auxillary systems $A_{1}^{\prime}$, $A_{2}^{\prime}$,....., $A_{n}^{\prime}$ to make it pure and then compute the minima of the sum over all $S_{A_{i} A_{i}^{\prime}}$ for $i=1,...,n$. 

The mathematical expression through which it is written in the following

\textbf{Definition:}

For a n- partite mixed state , with density matrix $\rho_{A_{1} A_{2} ..... A_{n}}$, the multipartite entanglement of purification is defined as,
\begin{equation}
    \Delta_{n(P)}(\rho_{A_{1} A_{2} ..... A_{n}}) = min_{|\psi\rangle_{A_{1}A_{1}^{\prime} A_{2}A_{2}^{\prime} ..... A_{n}A_{n}^{\prime}}} \sum_{i=1}^{n} S_{A_{i} A_{i}^{\prime}}.
\end{equation}

This boils down to the definition of bipartite entanglement of purification once n is taken to be 2 with appropriate normalization ($\frac{1}{n}$ factor in the above definition). Of course, for the bipartite case, the EoP is symmetric under the two parts ($S_{A A^{\prime}}=S_{B B^{\prime}}$). Let us call the bipartite EoP as $E_{P}(= \Delta_{2(P)})$.

\textbf{Properties:}

\textbf{1.} If one of the systems gets decoupled, $\rho_{A_{1}...A_{n}}=\rho_{A_{1}...A_{n-1}} \otimes \rho_{A_{n}}$,then
\begin{equation}
    \Delta_{P}(A_{1}:...:A_{n}) = \Delta_{P}(A_{1}:...:A_{n-1}).
\end{equation}

\textbf{2.} For a n-partite pure state $|\psi\rangle _{A_{1}...A_{n}}$,
\begin{equation}
    \Delta_{P}(A_{1}:...:A_{n}) = \sum_{i=1}^{n} S_{A_{i}}.
\end{equation}

\textbf{3.} For a n-partite product state ($\rho_{A_{1}...A_{n}}=\rho_{A_{1}}\otimes\rho_{A_{2}} \otimes.... \otimes \rho_{A_{n}}$), 
\begin{equation}
    \Delta_{P}(A_{1}:...:A_{n}) = 0.
\end{equation}

\textbf{4.} $\Delta_{p}$ is bounded from above as follows,
\begin{equation}
    \Delta_{P}(A_{1}:...:A_{n}) \leq min_{i} \left(S_{A_{1}} +...+ S_{A_{1}...A_{i-1}A_{i+1}...A_{n}} + ...S_{A_{n}}\right).
\end{equation}

\textbf{5.} $\Delta_{p}$ is bounded from below as follows,

\begin{equation}
    \Delta_{P}(A_{1}:...:A_{n}) \geq I(A_{1}:...:A_{n}),
\end{equation}
where $I(A_{1}:...:A_{n})$ is the n-partite mutual information.

All of these are followed by bipartite EoP as well once one takes $n=2$.

\subsection{Holographic Duals:}

\begin{figure}[t]
\centering
\includegraphics[width=0.49\textwidth]{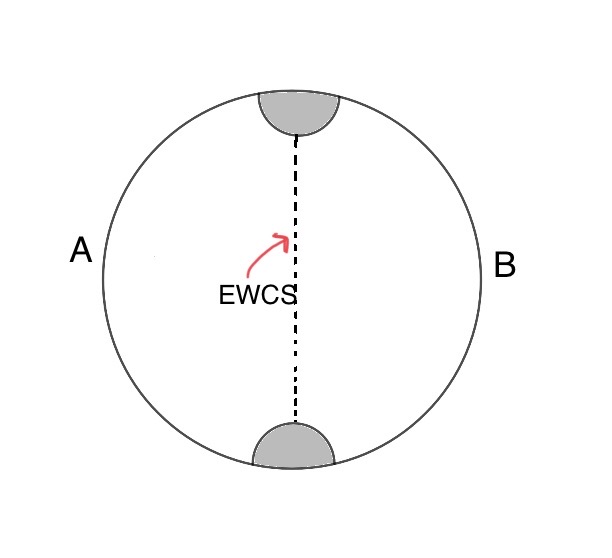}
\caption{Entanglement Wedge Cross Section (holographic dual of Bipartite Entanglement of Purification)}
\label{fig:2}
\end{figure}

The holographic duals of various purification measures have been proposed in various articles \cite{Takayanagi:2017knl, Umemoto:2018jpc, Jokela:2019ebz, Tamaoka:2018ned}.
Here we firstly discuss the holographic dual ($E_{W}$) of the bipartite EoP ($E_{P}$) and then the n-partite case. For holographic states, it was conjectured that the holographic dual of $E_{P}$ is the minimum entanglement wedge cross-section which is the dotted line in figure \ref{fig:2}. 
The mathematical definition of entanglement wedge cross section is the following ,

$ E_{W}(A : B) = min \{ Area(\Gamma) ; \Gamma \subset M_{AB} - M_{A\cap B}\}$ separates  $A\setminus B$ and $B\setminus A $ ,

\textit{where $M$ denotes the entanglement wedge of some specified interval in the boundary CFT. $A/B = \left(A - A\cap B\right)$ and $B/A= \left(B - B\cap A\right)$.} 
\cite{Takayanagi:2017knl, Umemoto:2018jpc, Bao:2018gck}

\begin{figure}[t]
\centering
\includegraphics[width=0.49\textwidth]{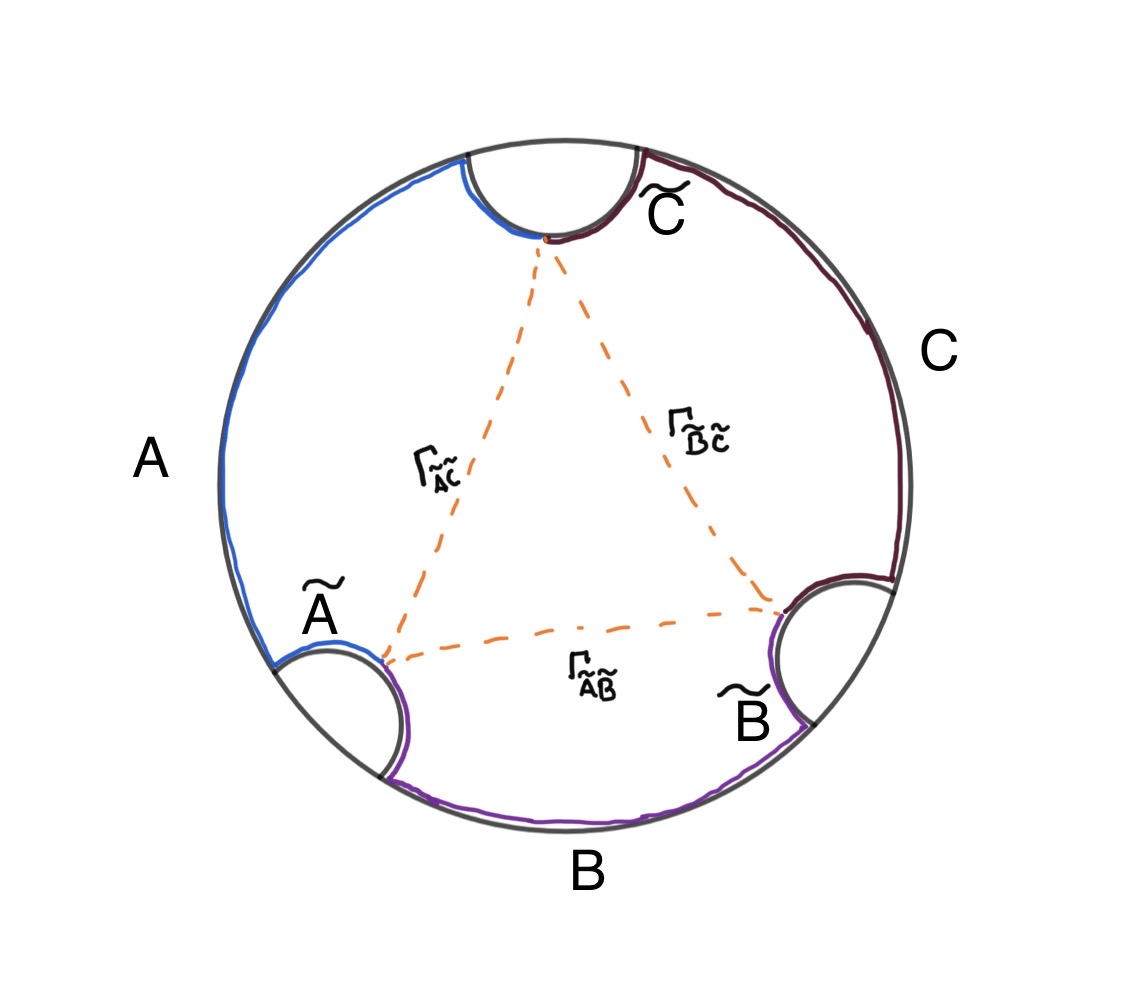}
\caption{EWCS for tripartite EoP: The combination of A, B, C and the the HRT geodesics are combined as $\Tilde{A}\Tilde{B}\Tilde{C}$ and considered to form a tripartite geometric pure state.}
\label{fig:3}
\end{figure}

For multipartite states, one needs to consider subregions involving boundary and bulk subregions to redefine $\Tilde{A}$, $\Tilde{B}$, $\Tilde{C}$ (for a tripartite case) and then compute the multipartite minimal entanglement wedge cross-section $\Gamma_{\Tilde{A}\Tilde{B}\Tilde{C}}$, where $\Tilde{A}\Tilde{B}\Tilde{C}$ is a geometric pure state. This is pictorially described in figure \ref{fig:3} \cite{Umemoto:2018jpc, Bao:2018fso}. Actually in case of bipartite entanglement of purification as well, the boundary subregions $A \cup B$ is typically a mixed state, but $\left( A \cup B \cup HRT_{1} \cup HRT_{2}\right)$ is considered to be a geometric pure state and the minimal length dividing the whole system into two is considered to be the bipartite entanglement of purification. The HRTs serve as the ancilla systems ($A_{i}^{\prime}$ parts mentioned in subsection \ref{PuriQM}.) added to make the geometric state a pure state.

In \cite{Takayanagi:2017knl}, it has been checked that $E_{W}$ follows the same set of properties as $E_{P}$, whereas, in \cite{Umemoto:2018jpc}, $\Delta_{n(W)}$ and $\Delta_{n(P)}$ have been found to share the same set of properties. 


\section{A Note on Page Curve Study and Toy Model:}{\label{section3}}

This is a very brief review of the recent program \cite{Almheiri:2019yqk, Almheiri:2019qdq, Almheiri:2019hni, Almheiri:2019psf, Akers:2019nfi} that has been instrumental in describing the time evolution of a black hole to be a unitary process by considering the combination of an evaporating black hole and the Hawking radiation to form a pure state. It has then been shown that using particular techniques, one can show that separately both system's entanglement entropy follows the same curve which is not ever-growing, but indeed comes down after Page time. This is a path-breaking result since it is the very first time that some program has been able to arguably solve the longstanding information paradox. Although it was always argued since the discovery of AdS/CFT that it can solve the information paradox, this is the very first concrete example where people have been able to show it in a somewhat convincing manner.

The technique through which this program was successful to achieve such a task is though yet to be made complete sense of. It involves the introduction of certain bulk regions called islands, which is essential to derive a Page curve for the emitted Hawking radiation along with new but familiar concepts of the quantum extremal surface. In the following subsections, we will first describe the basic ideas of the computations and then discuss a simple toy model involving multiboundary wormhole which successfully reproduces the Page curve and gives us an interesting and intuitive understanding of the actual computations just by considering classical RT surfaces instead of quantum extremal surfaces. 

\subsection{Information Paradox and Resolutions (Islands) :}
Information is ideally considered to be a sacred thing, which one should always keep track of. If there is information flow between two parts of the system, then the information missing in one part should necessarily show up in the other part. But, in the case of the black hole, it has been a longstanding problem in such a scenario. In the case of black holes, the way one typically compares information inside and outside is by specifying the entanglement between the two systems. The paradox appearing in this computation had been a peculiar one since the information found in the radiation outside the black hole seemed to be more than what the black hole could store. If we talk in terms of entanglement entropy, which is a standard information measure between two entangled states, the entanglement entropy of the radiation outside the black hole was found to be growing over for a very long time, whereas the black hole's entropy seemed to become less and less over time. They crossed each other much before the radiation entropy saturates. But, since the evaporating black hole and the radiation states should form a combination which is a pure state, the entanglement between them for all times should be the same. This is one of many ways in which the information paradox can be realized. Let us explain a bit more concretely.

Let us assume that the radiation state is considered just combinations of the Hawking quantas radiated by an evaporating black hole. For all such Hawking quantas, there is a partner quanta behind the black hole horizon, which is entangled to its outside partner. Let us schematically write the radiation quantas and their partner modes as a combination, which is a pure state, 

\begin{equation}
   | \psi \rangle_{rad} \equiv \sum_{\omega, n} e^{-\frac{\omega n}{2}} | n \rangle_{in} | n \rangle_{out}
\end{equation}
at any point of time t, where the time is kept track of the frequencies $\omega$ (summed over) and n counts entangled pair of Hawking quantas emitted till that time. 

Now, tracing over $| n \rangle_{in}$ states, one can find out the reduced density matrix of the out-state ($\rho_{rad, out}$), which in this case comes out to be in form of a thermal density matrix,

\begin{equation}
    \rho_{rad, out} \equiv \sum_{\omega, n} e^{-\omega n}  | n \rangle_{out} \langle n |_{out}.
\end{equation}

This leads to the paradox since once the entanglement entropy is calculated for this reduced density matrix following the usual formula of von Neumann entropy, it keeps growing until the black hole evaporates (number of n increases). On the other hand, the Bekenstein Hawking entropy, which is supposed to be representative of black hole entropy keeps decreasing as the black hole evaporates and the area decreases. After some time (known as the Page time),

\begin{equation}
    S(\rho_{rad,out}) > \frac{A_{BH}}{4G}.
\end{equation}

The situation then complicates as the bipartite entangled state between the radiation outside and the black hole becomes oversaturating what the black hole can entangle with. In another way, the black hole is entitled to more entanglement than it has microstates available. 

\begin{figure}[t]
\centering
\includegraphics[width=0.80\textwidth]{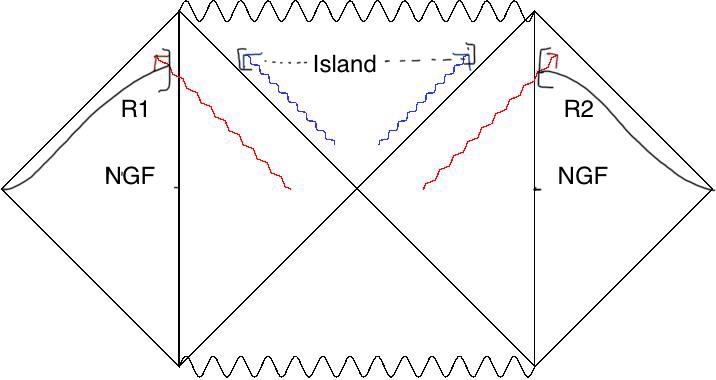}
\caption{Penrose diagram of two sided Black Hole with nontrivial island included (red and blue lines represent Hawking partner modes outside and inside the black hole horizon respectively.)}
\label{fig:4}
\end{figure}

There have been several attempts and effort in solving this paradox and deriving a formula which follows the well-known Page evolution (grows initially, then comes down to zero after Page time). But, until very recently, there has not been a convincing way in which people have been able to do it. In these recent set of papers \cite{Almheiri:2019yqk, Almheiri:2019qdq, Almheiri:2019hni, Almheiri:2019psf, Akers:2019nfi}, the resolution is brought in by introducing certain regions, termed as islands, which are null before Page time, but are non-null region (behind the black hole horizon) after the Page time. This also has to be considered for computing the entanglement entropy of the radiation states. Given such a situation, one has to also work with the idea of quantum extremal surfaces since while computing the entanglement associated with the nontrivial regions behind the horizon, Ryu-Takayanagi also contributes in a nontrivial way to the entanglement entropy of the radiation states.

Let us now look at \ref{fig:4} to get a better understanding. We consider a two-sided BH in AdS$_{3}$. The extreme left and extreme right regions represent non-gravitational flat space (NGF) coupled with the asymptotic AdS boundaries. These are needed as we are considering evaporating black holes and these coupled NGFs provide us a way to introduce absorbing boundary conditions in the shared boundary through which outside Hawking quantas can escape (unlike eternal BH case, where the outside quantas are reflected from the AdS boundary to feed the black hole back). Using this, we compute the entanglement entropy of the outside quantas in the NGFs. It is like stacking up the quantas escaping AdS in the NGFs. But simply doing these would again lead to the usual paradox. Say we compute entanglement entropy at an anchored time-slice t for a region from infinity (in the NGF) to very near the AdS boundary on both sides of AdS. Let us call these two regions R1 and R2 and their union ($R1 \cup R2 = $) R. $S_{out}[R]$  would again grow for a very large time and lead to the information loss. 

The introduction of the islands comes to the rescue here along with the consideration of quantum extremal surfaces. In the picture, where islands are included (shaded region behind the horizon, after the Page time), the new notion of entanglement entropy for the outside quantas looks like,

\begin{equation}
    S_{out}[R](new) =\genfrac{}{}{0pt}{5}{\large{min}}{I} \left[ \genfrac{}{}{0pt}{5}{\large{ext}}{I}\left\{ \frac{A(\partial I)}{4 G} + S_{usual}[R\cup I] \right\} \right],
\end{equation}

 where $\partial I$ is the boundary of the region enclosed by the islands. 
 \begin{figure}[t]
\centering
\includegraphics[width=0.70\textwidth]{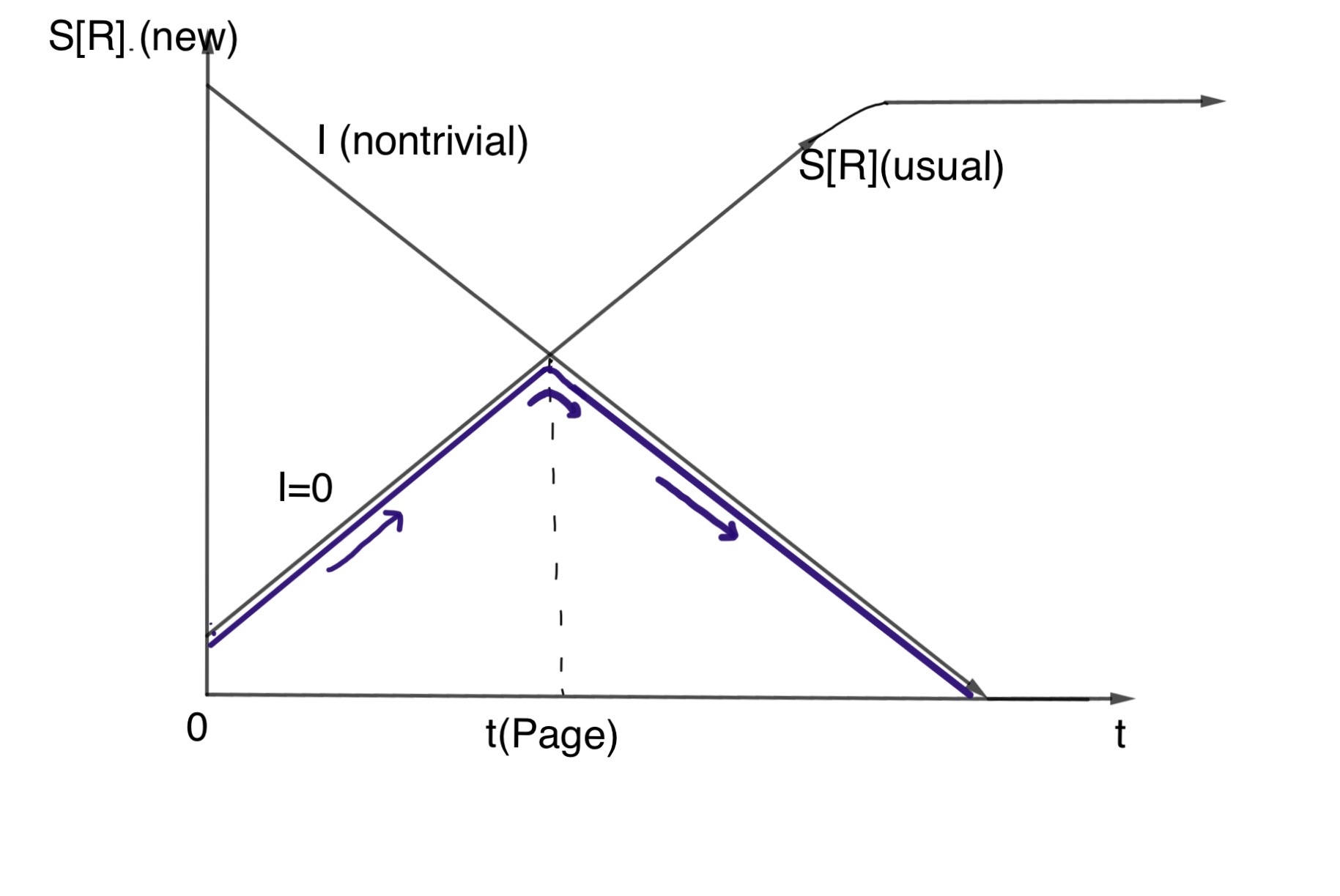}
\caption{Choice of islands before and after Page time and Page Curve}
\label{fig:5}
\end{figure}
 
 For a given timescale, firstly one has to take all choices of I (any interval in AdS$_{3}$, inside or outside the horizon can be a candidate for I). Then the sum of the two things in the curly bracket has to be extremized. The notion would be that in general, there exist more than one choice of I for which the sum is extremized. One has to choose the one which minimizes the sum at any given time. This solves the paradox since one finds that before Page time, minimal choice of island is the null (trivial) one and therefore up to that time, the new entropy is the same as the usual one which grows. In this case, $\frac{A(\partial I)}{4 G}$ is zero whereas $S_{new}= S_{usual}$. But after the Page time, the choice of the island which minimizes the sum among other choices of extremas is the one just behind the horizon. In that case, $\frac{A(\partial I)}{4 G}$ becomes the dominant contributor, as in the other part, both the entangled Hawking quantas (inside and outside black hole) are included. In that case, this piece contributes much less as the Hawking quantas are purified. This is an important point we would come back to while making connections to multipartite entanglement of purification. But, the dominant contributor ($\frac{A(\partial I)}{4 G}$) decreases over time which helps in the production of a Page curve (see Figure \ref{fig:5}).

\subsection{Multiboundary Wormholes and a simple toy model:}

In this subsection, we briefly discuss multiboundary wormholes in AdS$_{3}$ and then we discuss the toy model introduced in \cite{Akers:2019nfi}, where the authors have shown that classical RT surfaces can also reproduce a Page curve in some situations and the aspects of the newly introduced islands can be given an intuitive understanding from the perspective of quantum error correction \cite{Harlow:2018fse, Pastawski:2015qua, Dong:2016eik}.

Multiboundary wormholes are situations where many boundary CFTs are connected by a wormhole. All these different boundaries are independent of each other. The construction of multiboundary wormholes in $AdS_{3}$ is a well-discussed topic, but an active area of research in itself. In usual understanding, multiboundary wormholes can be thought of as multiple exits created by quotienting AdS$_{3}$ and by removing semicircles from a timeslice of pure AdS$_{3}$ by orientation reversing isometries in the upper half plane. This defines the fundamental domain. Since in three spacetime dimensions, true dynamical degrees of freedom are lacking, only global topological data and boundary dynamics classify a classical saddle implying that for smooth asymptotically AdS$_3$, all geometries locally belong to the same universal class and are distinguished only by global features. 

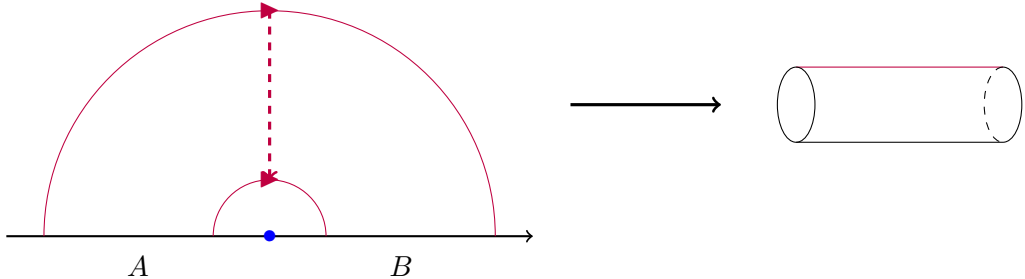
\begin{figure}[t]
	\centering
	\begin{tikzpicture}[scale=1.0]
	\draw[->,thick] (-3.5,0) to (3.5,0);
	\draw[-,color=purple] (-0.75,0) arc (180:0:0.75);
	\draw[-,color=purple] (3,0) arc (0:180:3);
	
	\node[color=purple,rotate=-90] at (0,0.75) {$\blacktriangle$};
	\node[color=purple,rotate=-90] at (0,3) {$\blacktriangle$};
	\draw[->,very thick,dashed,purple] (0,3) to (0,0.75);
	\draw[->,very thick] (4,1.75) to (6,1.75);
	
	\draw[-,color=purple] (7,2.25) to (9.75,2.25);
	\draw[-] (7,1.25) to (9.75,1.25);
	\draw[-] (7,1.75) ellipse (0.25 and 0.5);
	\draw[-] (9.75,2.25) arc (90:-90:0.25 and 0.5);
	\draw[-,dashed] (9.75,2.25) arc (90:270:0.25 and 0.5);
	
	\node[color=blue] at (0,0) {$\bullet$};
	\node[color=black] at (-1.75,-0.4) {$A$};
	\node[color=black] at (1.75,-0.4) {$B$};
	\end{tikzpicture}
	\caption{Two boundary case and Horizon length, equivalent to EWCS for bipartite system. }
	\label{fig:6}
\end{figure}

In AdS/CFT, this is related to the study of n fold tensor product of CFT states in different boundaries. For $n=2$, the resulting geometry is of a BTZ which is dual to a TFD (thermofield double) state.  \cite{Caceres:2019giy} is a recent paper that discusses these things in detail. Figure \ref{fig:6} is the way one creates two boundaries by removing two semicircles from pure AdS$_{3}$ slice at $t=0$ through a killing vector that generates dilatation. The standard way of addressing dynamical questions in CFTs is the formalism known as Schwinger-Keldysh, which in context of holography is translated as considering multiboundary geometries in Euclidean and Lorentzian signature and glueing across a surface of zero extrinsic curvature (boundary anchored geodesics). As spacelike slice of AdS$_3$ always maps on to the Poincare disk by stereographic projection, we start with a Poincare disk and take quotient by a single hyperbolic isometry producing a Riemannian surface with constant -ve curvature everywhere. This manifold is the one that one gets if one cuts a strip bounded by geodesics anchored on the boundary out of the disk and glues it shut. This produces a time-symmetric slice at $t=0$ of a two sided BTZ. Let us give a surprise at this point. Figure \ref{fig:6} is the fundamental domain of a 2 sided BTZ, defined by removing two semicircles that are related through a dilatation. This in actual would look like \ref{fig:2}, where A and B are the two boundaries where two CFTs live. We will discuss these connections in detail in the next section.

Nevertheless, one can introduce more and more exits by removing more and more semicircles in an orientation reversing way (again, discussed in Appendix \ref{app A}) on one side of the smaller semicircle of Figure \ref{fig:6}. These removals simply correspond to quotienting by more and more number of isometries. Removing semicircles from the other side would mean introducing handles. \footnote{ The introduction of a handle involves removing two semicircles from two sides of the lower semicircle of Figure \ref{fig:6}, but it also reduces the number of exits/horizons by one. }

 A multiboundary wormhole can be understood as a diagram that resembles a pant with leg space more than two. In such a construction, all the different horizon lengths can be tuned or changed independently in terms of the parameter in the timeslice of AdS$_{3}$ through which the semicircles are removed. But for a two-boundary case, there is only one horizon, which both the CFT sees and there is only one parameter involved which is the ratio of the radius of the semicircles in figure \ref{fig:6}. Starting from $n>2$, an n-boundary wormhole would have parameters such that all the horizons can be made big or small independently using a non-overlapping set of parameters.
 
                       Using the above-mentioned fact, the authors in \cite{Akers:2019nfi} introduced a toy model, in which they considered a multiboundary wormhole where one of the horizons is much much longer to start with than all other horizons. They begin from $n=3$ case (thus with the independence of parameters are allowed from the very start). They describe the boundary of the bigger horizon to be the evaporating black hole, whereas other black holes, which are of much much smaller in size and live in the smaller exits of the MBW picture, are considered to be the Hawking quantas radiated. With each emission of Hawking quanta, a new exit is created and the horizon length of the bigger evaporating black hole is reduced very little.
                       The authors neglect the discussion regarding new topology (exits) created in such a process and any issue with bulk dynamics. 
                       
         \begin{figure}[t]
	     \centering
	     \includegraphics[width=0.40\textwidth]{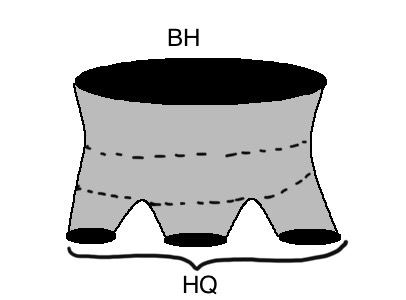}
	     \includegraphics[width=0.40\textwidth]{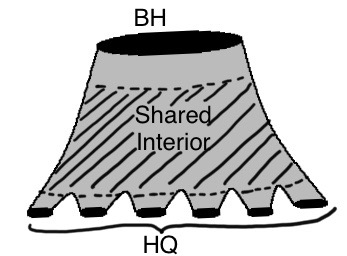}
	     \caption{Pantlike diagrams of MBW and the shared interior }
	     \label{fig:7}
         \end{figure}

         An important assumption the authors make is that the ADM energy is conserved during the evaporation process. Using this fact, one can define the smaller black holes (dual to emitted Hawking quantas in such a model) to be of comparable length scale (the horizon length of all of them made almost equal, independently) which is much much less than the evaporating one. But as along the process, the ADM energy conservation is used and the evaporating black hole reduces in size, the HRT surfaces of the union of the disconnected Hawking quantas choose different horizons in different times \footnote{in this picture, both the horizon corresponding to bigger black hole and union of the horizons of the smaller ones are homologous to both the bigger black hole as well as the union of smaller black holes/Hawking quantas.}. Primarily the smaller choice of HRT surface is the union of disconnected horizons of the smaller exits (/Hawking quantas) whereas, in a later time, where the number of emitted quantas is much large, the minimal HRT choice is the horizon of the evaporating black hole. This study produces a Page-like curve since in later times, a different HRT surface is automatically more favorable and the region between the bigger and smaller horizons is intuitively understood as the nontrivial island in the later times. 
         
         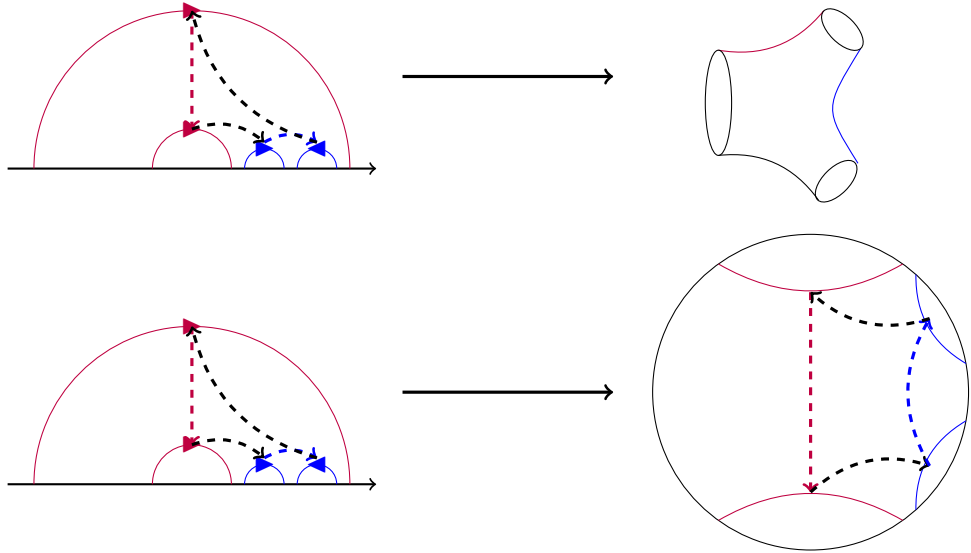
\begin{figure}
	\centering
	\begin{tikzpicture}[scale=0.7]
	\draw[->,thick] (-3.5,0) to (3.5,0);
	\draw[-,color=blue] (1,0) arc (180:0:0.75/2);
	\draw[-,color=blue] (2.75,0) arc (0:180:0.75/2);
	
	\draw[-,color=purple] (0.75,0) arc (0:180:0.75);
	\draw[-,color=purple] (3,0) arc (0:180:3);
	
	\node[color=blue,rotate=-90] at (1+0.75/2,0.75/2) {$\blacktriangle$};
	\node[color=blue,rotate=90] at (2.75-0.75/2,0.75/2) {$\blacktriangle$};
	
	\node[color=purple,rotate=-90] at (0,0.75) {$\blacktriangle$};
	\node[color=purple,rotate=-90] at (0,3) {$\blacktriangle$};
	
	\draw[->,very thick,dashed,purple] (0,3) to (0,0.75);
	\draw[->,very thick,dashed,blue] (1+0.75/2,0.5) to[bend left] (2.75-0.75/2,0.5);
	\draw[->,very thick,dashed,black] (0,0.75) to[bend left] (1+0.75/2,0.5);
	\draw[->,very thick,dashed,black] (2.75-0.75/2,0.5) to[bend left] (0,3);

	\draw[->,very thick] (4,1.75) to (8,1.75);

	\draw[-,color=purple] (3+7,2.25) to[bend right] (3+9,3);
	\draw[-,color=blue] (3+9.7,2.28) .. controls (3+9,1.14) .. (3+9.65,0.1);
	
	\draw[-] (3+7,1.25) ellipse (0.25 and 1);
	\draw[-] (3+7,0.25) to[bend left] (3+9-0.1,-0.6);
	
	\draw[-,rotate around={45:(3+9,2.5)}] (3+9.35,2.35) ellipse (0.25 and 0.5);
	\draw[-,rotate around={135:(3+9.2,-0.7)}] (3+9.5,-1.05) ellipse (0.25 and 0.5);

	\draw[->,thick] (-3.5,-6) to (3.5,-6);
	\draw[-,color=blue] (1,-6) arc (180:0:0.75/2);
	\draw[-,color=blue] (2.75,-6) arc (0:180:0.75/2);
	
	\draw[-,color=purple] (0.75,-6) arc (0:180:0.75);
	\draw[-,color=purple] (3,-6) arc (0:180:3);
	
	\node[color=blue,rotate=-90] at (1+0.75/2,-6+0.75/2) {$\blacktriangle$};
	\node[color=blue,rotate=90] at (2.75-0.75/2,-6+0.75/2) {$\blacktriangle$};
	
	\node[color=purple,rotate=-90] at (0,-6+0.75) {$\blacktriangle$};
	\node[color=purple,rotate=-90] at (0,-6+3) {$\blacktriangle$};
	
	\draw[->,very thick,dashed,purple] (0,-6+3) to (0,-6+0.75);
	\draw[->,very thick,dashed,blue] (1+0.75/2,-6+0.5) to[bend left] (2.75-0.75/2,-6+0.5);
	\draw[->,very thick,dashed,black] (0,-6+0.75) to[bend left] (1+0.75/2,-6+0.5);
	\draw[->,very thick,dashed,black] (2.75-0.75/2,-6+0.5) to[bend left] (0,-6+3);

	\draw[->,very thick] (4,-6+1.75) to (8,-6+1.75);

	\draw[-,color=black] (14.75,-6+1.75) arc (0:360:3);

	\draw[-,color=purple] (10,-6.6867) to[bend left] (13.50,-6.6867);
	\draw[-,color=purple] (10,-1.8133) to[bend right] (13.50,-1.8133);
	
	\draw[-,color=blue] (13.75,-6.48607) to[bend left] (14.7,-4.79544);
	\draw[-,color=blue] (13.75,-2.01393) to[bend right] (14.7,-3.70456);
	
	\draw[->,very thick,dashed,purple] (11.75,-4.25+1.90) to (11.75,-4.25-1.90);
	\draw[->,very thick,dashed,black] (11.75,-4.25-1.90) to[bend left] (14.01, -5.65);
	\draw[->,very thick,dashed,blue] (14.01, -5.65) to[bend left] (14.01, -2.85);
	\draw[->,very thick,dashed,black] (14.01, -2.85) to[bend left] (11.75,-4.25+1.90);

	\end{tikzpicture}
	\caption{The three-boundary Riemann surface as quotients of the two-boundary Riemann surface. The three-boundary surface is obtained by pinching one of the boundaries into two. The island is marked by the closed region spotted by the dotted purple, black, blue and black line respectively. }
	\label{figs:tikz1}
\end{figure}
         
\begin{figure}[t]
         \centering
         \includegraphics[width=0.50\textwidth]{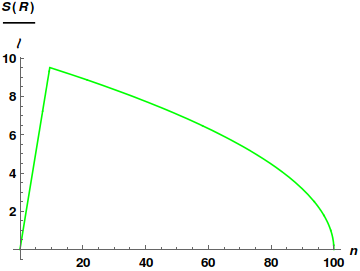}
         \caption{Page Curve from the MBW toy model}
         \label{fig:8}
         \end{figure}

         The relation between entropy and ADM energy in $AdS_{3}$ is the following
         \begin{equation}
             S = 2 \pi \sqrt{\frac{c E}{3}}.
         \end{equation}
         Now, in three bulk spacetime dimensions, the area of the HRT surface is simply the length. Let us consider the initial length of the horizon of the evaporating black hole is $L_{0}$ which decreases over time as it emits more and number of smaller black holes with horizon lengths $\ell$. Using the relation between length (entanglement entropy) and ADM energy,
         we can show that at any point of time, where n smaller black holes have been emitted, the horizon length of the bigger black hole reduces in the following way,
         \begin{equation}
             L(BH)= \sqrt{L_{0}^{2} - n \ell^{2}}
         \end{equation}
         whereas the union of the length of the horizons of the smaller horizons scales like $L(HQ)= n \ell$ \footnote{$L_{HQ}$ corresponds to the length of the horizons of the smaller black holes that are analogs of Hawking quantas in the radiation}. For smaller values of n, $L_{HQ}$ is the minimal choice, which grows over time as n increases. The $L_{BH}$ decreases as time moves forward. At certain timescale $n \sim \frac{L_{0}}{\ell}$, $L_{BH}$ and $L_{HQ}$ become comparable and after that $L_{BH}$ becomes the minimal HRT choice. This is the way one can get a Page-like curve through this toy model. (see figure \ref{fig:8})
         
         This region between the chosen HRTs at different times (here the number of exits, n, is considered to be the analog of time), has also similarities with the shared interior that appears in a study of quantum error correction through bulk reconstruction picture. Through this similarity, the authors in \cite{Akers:2019nfi} provided a possible understanding of the islands, which is due to the full and restricted set of observables that can be reconstructed depending upon which surface is chosen. According to them, as the shared interior is not dual to any single boundary subregion, it appears as a quantum error-correcting region in the computation of entanglement entropy. 
         Now, in the next section, we discuss the lessons we learn from the two previous studies and make a few connections among them with the hope that we would get a better understanding of what the toy model implies and how can it be connected to the study of multipartite entanglement of purification.
 
 \section{Connections between Purification and MBW Toy Model:}{\label{section4}} 
 In this section, we firstly make the two pictures clearer by making several connections between them. Then we also discuss how the naive application of the island formula can lead to the earlier paradox. We also take lessons from other works that help us to get rid of the problems and make our understanding stronger in both the studies.

 \subsection{Connections to be drawn:}
 The two topics discussed in the two previous sections have striking similarities which are yet to be pointed out. We make following observations and constructions that help us understand the connections and also gives us a few important lessons that should be kept in mind while comparing the two scenarios.

 1. As we have already pointed out once in the previous section, the pictures of bipartite entanglement of purification where the bipartition made by choosing two disjoint, but substantially larger subregions of a timeslice of pure AdS$_{3}$ is very similar to the construction of a wormhole connecting two boundaries, where the boundaries exactly correspond to subregions A and B of the bipartite system. 
 
 To be more exact , in case of two boundaries, one takes two boundary anchored geodesics in Poincare disk, the fundamental domain (corresponding to the HRT surfaces of the region $A \cup B$ not sharing any endpoints) and uses a unique isometry (dilatation to be precise) that defines a bijective map from points on one of the geodesics to the closest points on the other. This map is the part where one identifies points on the two geodesics periodically and glues them. This isometry doesn't involve any fixed points in the strip between the two geodesics. \footnote{Multiboundary wormhole constructions can involve isometries including pathologies like fixed points and closed timelike curves in general. But both can be avoided by making suitable choices as mentioned in \cite{Caceres:2019giy}. }
 
 2. Another striking similarity is the entanglement wedge cross-section for a bipartite state is the only possible horizon length that one can compute in a two boundary case. After the identification and glueing procedure is done among the two boundary anchored geodesics in the Poincare disk, all one needs to specify the two sided BTZ is not two, but a single geodesic specifying the horizon length.

 Now, we can also simply go on to pictures including more exits and compare the two cases in one of which, we increase the number of exits in the MBW picture and in the other picture, we introduce more number of disjoint subregions in one of the subregions A and B. \footnote{If we introduce more subregions in both sides, that would mean introducing handles in the MBW picture. We avoid such scenarios for the time being.} But there is one subtlety involved that one should keep in mind while doing so. For $n\geq 3$, one needs to remove two semicircles for introducing each new exit. Thus after $n=2$, in the partitioning of the boundary, we have to introduce two disjoint subregions at each step in the purification picture, a combination of which will be equivalent to the newly constructed boundary. 
 
 \begin{figure}[t]
\centering
\includegraphics[width=0.50\textwidth]{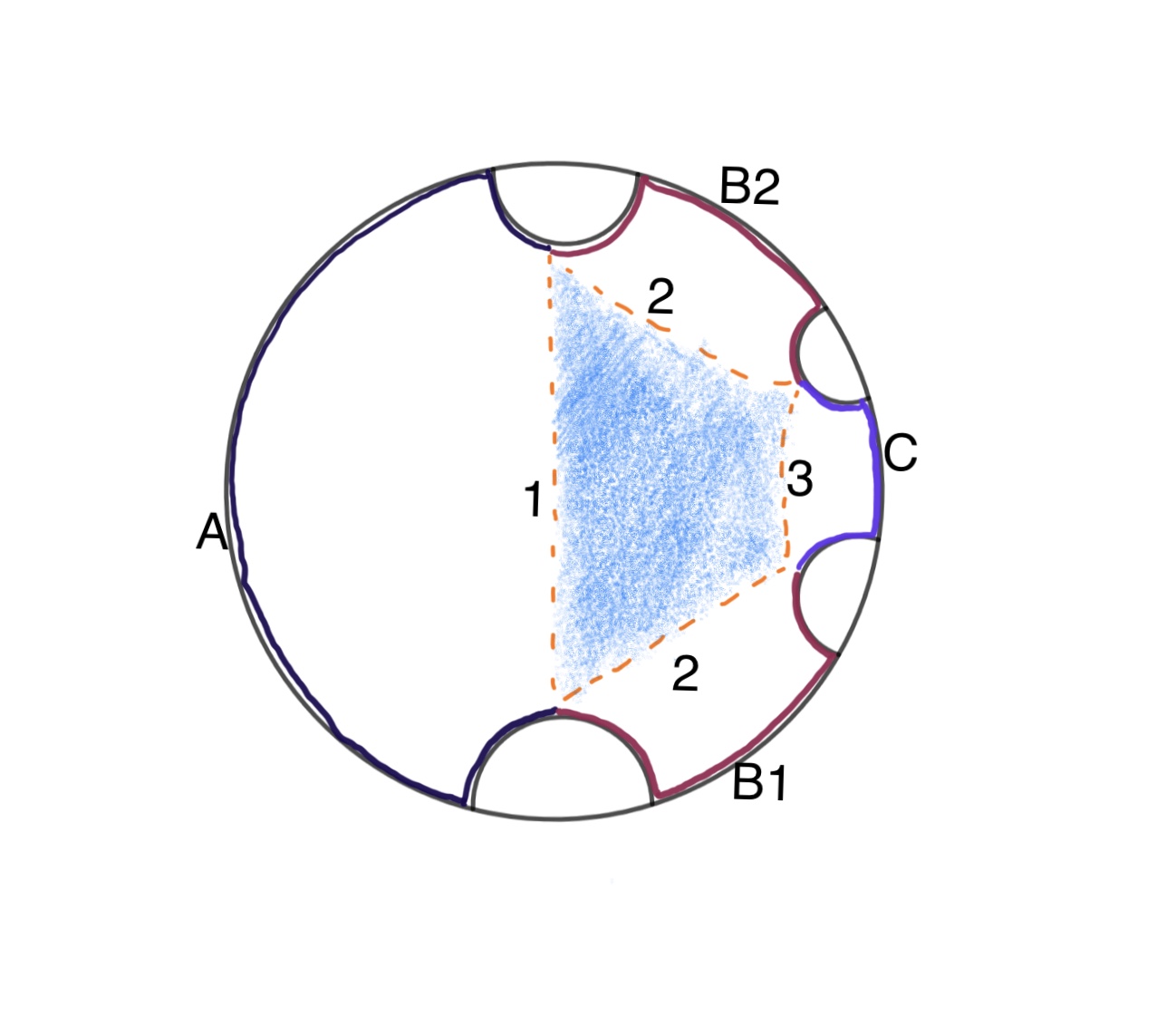}
\includegraphics[width=0.45\textwidth]{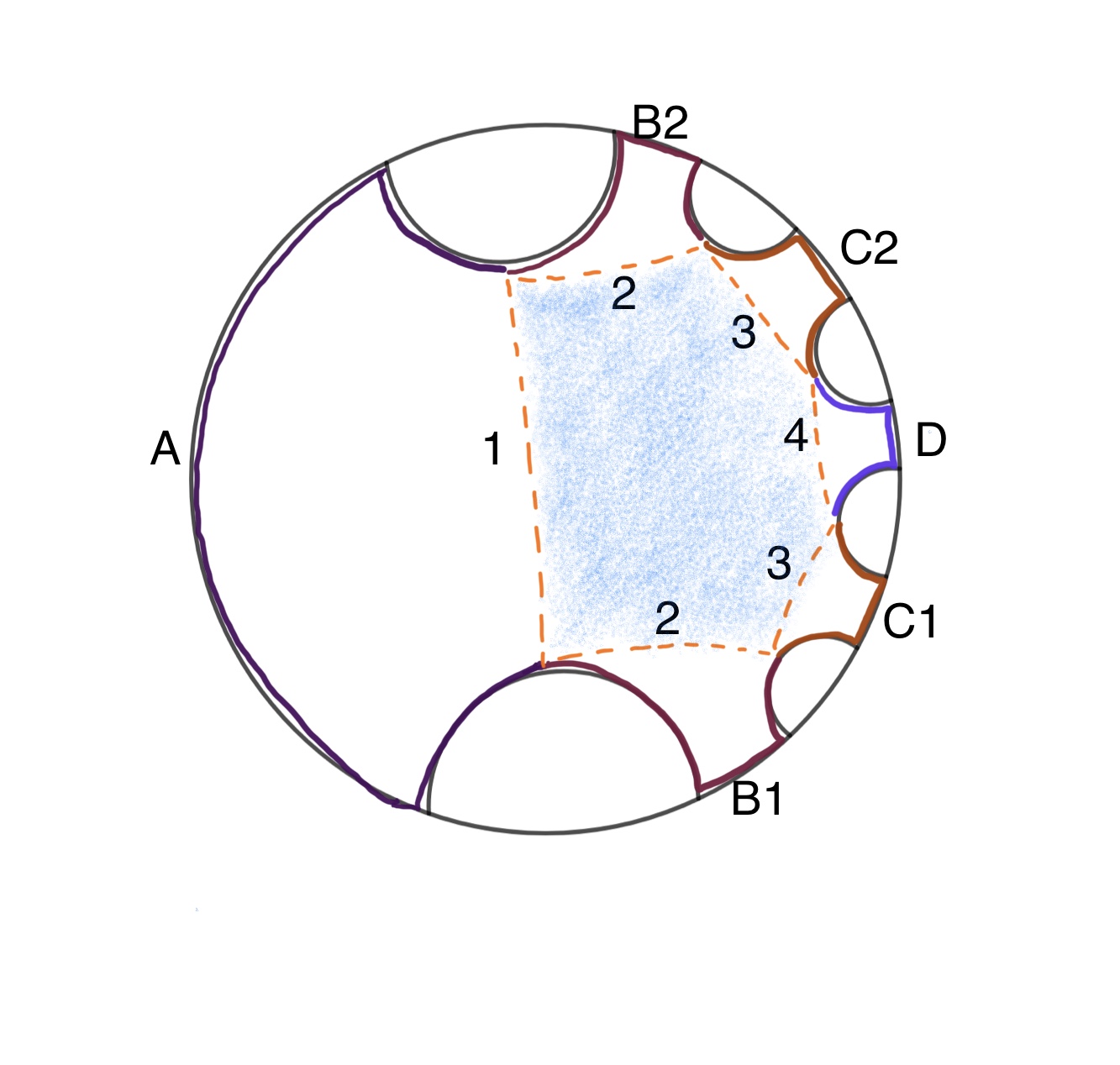}
\caption{Three and Four Boundary cases : Semicircles to be removed are  are marked. Blue-shaded regions represent the shared interiors. These are the choices that minimizes the boundary of the shared interior with respect to the corresponding geometric pure state.}
\label{fig:9}
\end{figure}

 In doing so, in each step, we should also keep decreasing the size of the other subsystem A very slowly so that our picture goes well with the previously introduced toy model. We also introduce new subsystems in such a way that their contributions in multipartite EoP are much smaller initially w.r.t the contribution of the bigger subregion A.
 
 We can define our newly defined subsystems in the following way so that it again goes well with the boundaries that are defined in the MBW case. We take two disjoint boundary intervals connected to A on two different sides and call them B1 and B2. We call their union to be a partition B. Similarly for more exits, we keep taking unions of intervals connected to B1 and B2 on two sides and define as a new partition. (as shown in figure \ref{fig:9})

3. Now, by looking at the two pictures, one can easily point out that the dotted lines in Figures 7 and 8 are equivalent to each other. But, in multipartite EoP, we take the sum of all of them, whereas, in the MBW toy model, they are treated as two different sets that naturally provide one with a way to choose one of them as the HRT surface. If we look at the figures more carefully, it is not hard to find out that the codimension 1 region enclosed by the multipartite entanglement wedge cross-section is the shared interior (/island) in the MBW picture. It is then obvious to define the multipartite entanglement wedge cross-section as the boundary of the nontrivial island. We thus get an understanding of the boundary of the nontrivial island after the Page time in terms of entanglement of purification between the evaporating black holes and the radiation quantas (union of all other boundary subregions except A).

4. An important fact while talking about the connections between the two scenarios is to realize the correspondence between a geometric pure state and the multiboundary wormholes connecting the bigger and the smaller black holes. In case of multipartite entanglement of purification, we consider the bulk HRT surfaces for the subsystems of the pure AdS along with the boundary subregions as a geometric pure state, \footnote{To be precise, the HRT surfaces of the multiboundary cases do not form a closed region by themselves. The shared interior is understood as the union of HRTs alongwith certain regions of the wormholes connecting different exits (See Appendix \ref{app B} for more details on this) (See Figure \ref{fig:7}). But since while considering a geometric pure state in the dual EoP picture, we make the choices of $\Gamma_{\Tilde{A}\Tilde{B}}$, $\Gamma_{\Tilde{B}\Tilde{C}}$ and $\Gamma_{\Tilde{A}\Tilde{C}} $ in such a way that the combination of them forms a minimally closed curve (See Figure \ref{fig:3}) among all other choices. This choice is always the minimal choice of choosing the boundary of the analog of the islands. (See Figure \ref{figs:tikz2})}  The multiboundary wormhole connecting CFTs at different exits act as a machine to make the whole multipartite state (/combination of multiple exits) a pure state. Making this connection helps us to understand the multiboundary wormhole along with the exits as a geometric pure state for which we consider the HRT surfaces to compute the entanglement entropy. \footnote{Some of these connections were also drawn and discussed in \cite{Bao:2018fso}. Indeed, their discussion is more detailed. But the goal of their work was to advertise the entanglement entropies of multiboundary wormholes as equivalent to entanglement of purification of pure AdS and advocate that entanglement entropies are easier to compute than the EoP. Our connections and study is a bit more safeguarded as we only make the connections by introducing subregion in a single side of the bigger subregion which doesn't introduce handles in the wormhole geometry.}

\subsection{Realization of Overcounting through Shared Interior: }

The previously mentioned comparisons and connections indeed support the connection between the islands and quantum error connection since multipartite EoP has well-discussed connections to quantum error corrections as well as discussed in \cite{Kudler-Flam:2018qjo}. But as mentioned in the formula of quantum extremal surface, if one computes the area (length in case of AdS$_{3}$) of the boundary of the island, it doesn't behave as per our expectation. This is because the $A(\partial I) = \Delta_{W}$ and it consists of both the union of smaller horizons as well as the bigger horizon. It also means that the length of the boundary of the nontrivial island would follow the properties followed by multipartite EoP, which we have already listed in section \ref{section2}.  

\begin{equation}
   A( \partial I) = \Delta_{W} = \sqrt{L_{0}^{2}- n \ell^{2}} + n \ell.
\end{equation}

\begin{figure}[t]
\centering
\includegraphics[width=0.48\textwidth]{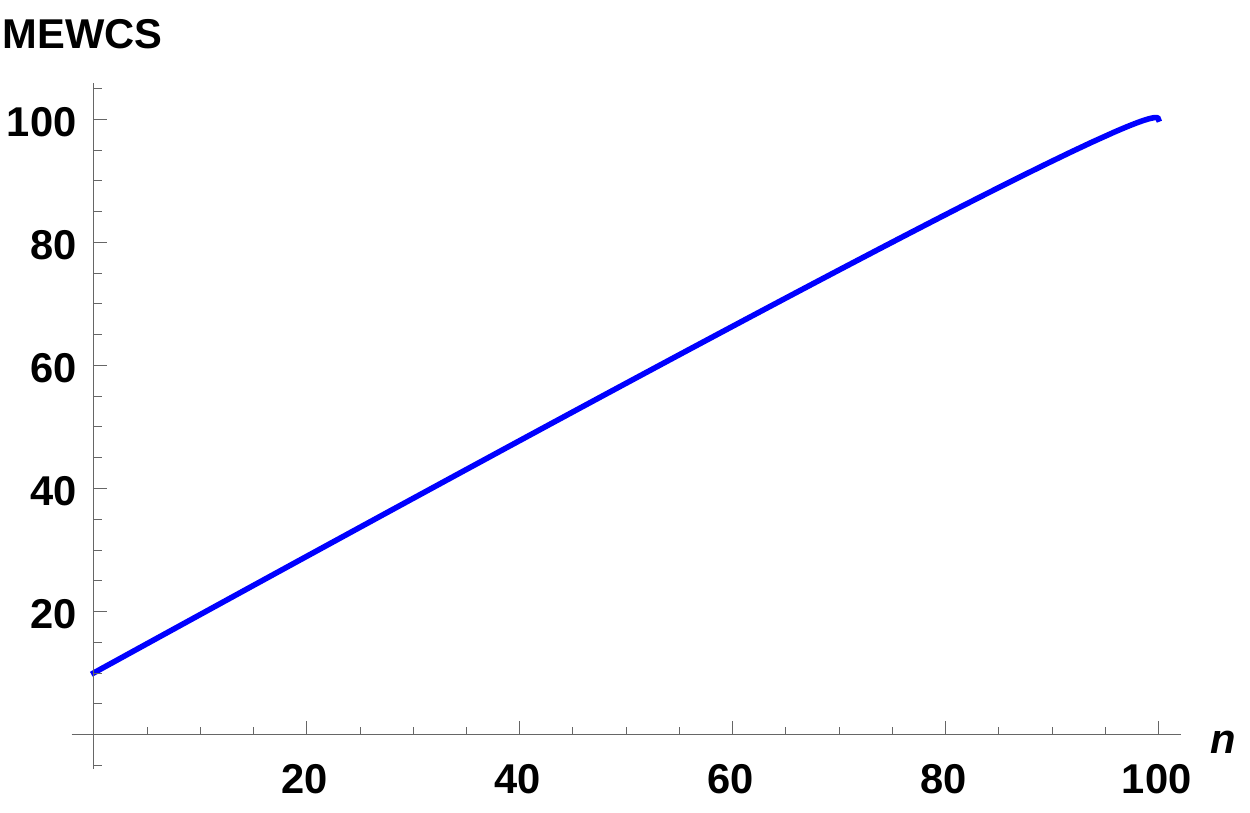}
\includegraphics[width=0.48\textwidth]{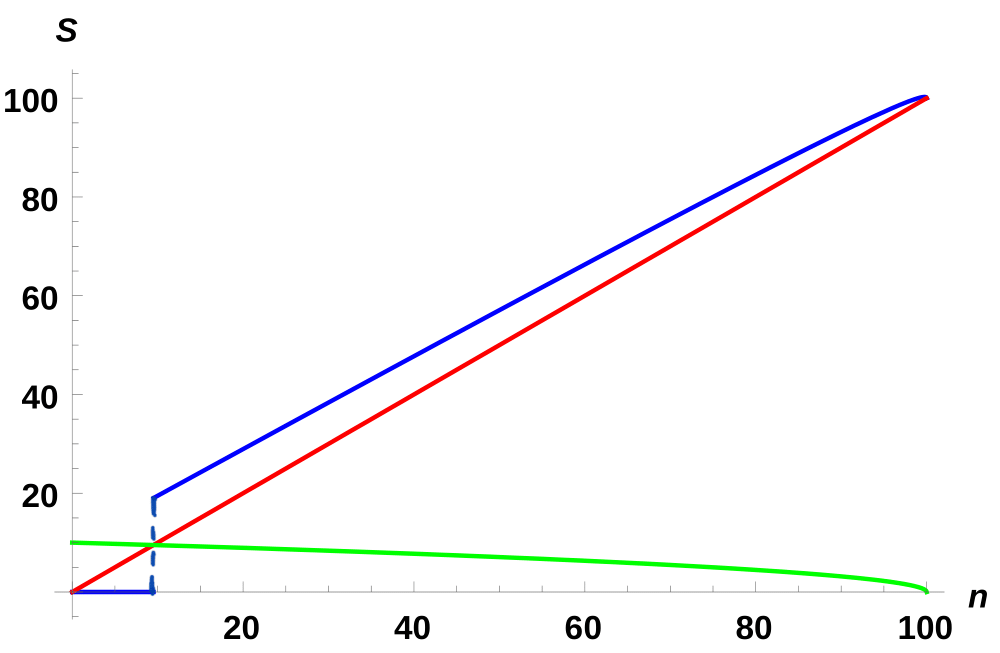}
\caption{(left) Growth of the multipartite EWCS (for the minimal choice). (right) Comparison between primary (red) and later (green) choice of HRT with the minimal island(blue) growth at different times.}
\label{fig:10}
\end{figure}

Now although the length of the bigger horizon keeps decreasing over time, the length of the combination of the smaller ones keeps increasing.  The the sum of them still grows (see figure \ref{fig:10}) until the black hole evaporates (in this case, this corresponds to the case where subregion A becomes so small that the entanglement wedge \cite{Harlow:2018fse} of the partitions simply become the union of the causal wedge of each of them). 

Therefore if one strictly assumes the shared interior to be the analog of the nontrivial island, its boundary area is evergrowing even after the Page time. (shown in figure \ref{fig:10}). Nevertheless, we prescribe the following resolution to the paradox. Our prescription is that this is again the same paradox that this whole program began to deal with. In the toy model, the authors try to realize the notion of the islands just through the classical HRT surfaces neglecting the bulk entropy part (second term in QES equation) assuming that the length of the smaller horizons individually is enough to keep track of the bulk entanglement entropy associated to the smaller black holes.

The argument is not so unsatisfying once we take into account that the analogs of the Hawking quantas are small black holes in the MBW picture, which are classical geometric objects. But from the point of entanglement of purification, when we consider the whole multipartite EWCS, we include both the bigger black hole horizon as well as the smaller ones. This, if translated to the statements made in \cite{Akers:2019nfi}, effectively means that we double count the bulk entropy of Hawking quantas in multipartite EWCS. Hence, once the island is included, in our calculations, the entanglement between the partner modes of the emitted quantas also contribute to the multipartite EoP. But, in fact, as the new HRT includes both the partners, they are purified. Multipartite EoP is insensitive to this purification and overcounts this to make the entanglement of purification larger than what it should be. 

The final resolution on the choice of multipartite entanglement wedge cross-section can be drawn from \cite{Balasubramanian:2014hda} in which again multipartite entanglement has been studied in details. Drawing connections from that paper, we can resolve the problem in the following way. As one of the black hole is considered to be much much bigger than the other ones, primarily all other horizon lengths can be considered as $\ell \rightarrow 0$. The reverse limit would be taking the smaller horizons $\ell$ to be finite whereas $L_{BH} \rightarrow \infty $. In both of these limits, the combined state behaves like a bipartite state \cite{Balasubramanian:2014hda} between the bigger BH and the union of the smaller ones. In that situation, the multipartite case boils down to a simplified bipartite case where the entanglement of purification reduces to usual entanglement entropy. Therefore, one can simply choose either the union of the horizons of smaller black holes or the horizon of the bigger black hole as the HRT surface depending on whichever is minimal at that time. But even in that scenario, if the newly entered shared interior is considered to be the analog of the nontrivial island, the growth of the length of the boundary of the island is paradoxical. The reason why this paradox arises only in the toy model is that in this case, the analog of islands is connected to the earlier null island. On the other hand, in case of the actual case, the island is behind the black hole horizon and is disconnected from the trivial island choice before the Page time.

If we treat different smaller black holes differently, we would have to necessarily consider multiparty entanglement of purification. Say we consider the bigger black hole as subsystem A, whereas n smaller black holes as $B_{1}$, $B_{2}$, ...., and $B_{n}$, then Multipartite EoP should be
\begin{equation}
    \Delta_{P} (A:B_{1}:....:B_{n}) = \frac{1}{n} min_{|\psi\rangle_{Pure}} \sum_{i=1}^{n} \left(S_{A A^{\prime}} + S_{B_{i} B_{i}^{\prime}}\right). 
\end{equation}
In this scenario, where all the smaller black holes are treated as a combination, and along with the bigger black hole, they form a pure state, we can take $A^{\prime}= B_{i}^{\prime}=\emptyset$ and therefore,
\begin{equation}
    \Delta_{(n+1)P} (A:B_{1}:....:B_{n})=\Delta_{(2)P} (A:B) = \frac{1}{2} (S_{A} + S_{B}).
\end{equation}
   
   Note that it is necessary to consider the full state as a bipartite pure state. A multi-partite pure state would not solve the problem. For example, if we considered that the combination of the big black hole and n smaller black holes to be an $(n+1)$-partite pure state, we would have to still apply the property 2 of the multipartite EoP. 
   
   \begin{equation}
       \Delta_{P} (A:B_{1}:....:B_{n}) =  \sum_{i=1}^{n} \left(S_{A } + S_{B_{i} }\right) = \sqrt{L_{0}^{2} - n \ell^{2}} + n \ell,
   \end{equation}

   where for each small black holes, even after the Page time, $\ell$ would be the HRT surface for individual smaller black holes instead of $L_{BH}$. Thus the multipartite EoP would still give us the ever-growing entanglement. Thus the resolution appears only when in the limit of a very large number of very small black holes, we take the union of the smaller black holes to be a single mixed state, which along with the large black hole state forms a bipartite pure state. 

Through this study, we learn important lessons about the two pictures which we write down in the following section.


\section{Discussion and Outlook:}{\label{section5}}
From the connections we made between multipartite entanglement of purification and multiboundary wormholes in AdS$_{3}$, we can take away following points.

1. First and foremost, the multipartite entanglement wedge cross-section represents the boundary of the islands described in the toy model of the evaporating black hole. We believe that knowing this would strengthen the possibility of building a concrete understanding of the islands as well as purification in several different ways e.g; quantum error correction, entanglement negativity and many more. Precisely, in the large n limit, i.e.; where the number of smaller exits is big, multipartite EoP and the shared island match completely. In the multiboundary wormhole picture, the shared island is the region behind all the horizons present (which is also the case in the original works discussing the actual model) and therefore it falls in the region known as entanglement shadow. Our work suggests that through the multiboundary wormhole construction, the entanglement shadow can have a description through the multipartite EoP of subregions in a vacuum AdS$_{3}$ slice. 

2. The reproduction of Page curve helps to describe an evaporating black hole as a unitary system since the Page curve is typically found in systems that evolve unitarily over time. Now, given the appearance of islands, or rather quantum error connection nakes sure that the unitarity of the black hole evaporation process is recovered,  one would hope that these two things are related. The natural way to somewhat realize the connection is of course the purification of the Hawking quantas after the Page time. This results in appearance of the nontrivial islands. In our study as well, we use ideas of purification regularly which give rise to the multipartite EWCS and the area enclosed is understood as the nontrivial island (quantum error correction). In other studies as well, for example , in \cite{Kudler-Flam:2018qjo}, people have explored connections between entanglement of purification and quantum error correction. But, it would be really interesting to understand such a connection as a triangular relation where the three vertices of the triangle correspond to unitarity, purification and quantum error correction \footnote{We thank Arnab Kundu for pointing out this interesting future direction.}. Regarding the line connecting unitarity to purification, a realization to start the study is the fact that a reduced density matrix (from which the purification is typically done) is derived by tracing out degrees of freedom from the initial pure state.

\begin{equation}
    \rho_{red} = Tr_{(pure-red)}[ \rho_{pure}].
\end{equation}

This tracing out is a non-unitary operation. Therefore the reduced density matrix indeed carries the effect of a non-unitary operation. Hence, it is not beyond expectation that to get back the unitarity completely one needs to apply purification to the reduced mixed state. This is a start which can be pursued in more details to get a better understanding of the above-mentioned triangle.

3. Although the islands can be intuitively understood as the shared interior, the length associated to the boundary of such a shared interior leads to a problem in overcounting due to which the entropy associated to the shared interior does not follow the Page curve (since the growth of the boundary of the nontrivial island after the Page time persists as it includes the previously chosen HRT surfaces as well).  To be precise, in the toy model, it is assumed that the RT surfaces take care of the bulk entropy between the fields that live on different sides of the HRT. But once the new choice of HRT is made, it has both the partner modes in there. The modes for which the partner modes are not yet inside the new HRT, their bulk entanglement with their partners is again taken care of by the new HRT. But if one computes the sum of the length of the shared interior simply considering it to be the boundary of the island, one again counts the bulk entanglement between the modes which have already been purified due to the choice of the new HRT. Let us call the shared interior SI. Then,

 \begin{equation}
     L(\partial(SI)) = L_{BH} + L_{HQ}.
 \end{equation}
 SI only comes into the picture after the choice of nontrivial island is minimal. Starting from that point, the boundary of SI also includes $L_{HQ}$, which have bulk entanglement between partner modes of the two sides of the previous HRT choice and this is how the overcounting can again come into the picture. 
 
 In \cite{Akers:2019nfi}, the authors introduce a second model involving handles and pairs of TFD states of the baby universe and the radiation states to understand the previous overcounting that in the first place led to the information paradox. We see here that even without introducing a new model, one can get the overcounting from the very first model by naively following the formula of the quantum extremal surface to include the whole length of the boundary of the nontrivial island and get back to the earlier paradox.

This is a warning that taking the intuitive understanding of islands too literally might lead to several problems. In this particular toy model, it is necessary to compute the lengths of the chosen HRTs only at any point in time. It is not only that one does not need to include the bulk entanglement, but it is also wrong to consider the remaining part of the quantum extremal surface formula in terms of the island. This subtlety might also capture important insights in making a connection between islands and QEC more concrete since QEC is well studied in the literature of EoP \cite{Kudler-Flam:2018qjo}.

4. For the problem of multipartite EoP, one can make simple calculations that show how the multipartite EoP grows over time and how different parts of it contribute to the Page curve in the multiboundary wormhole model of black hole evaporation in AdS$_{3}$. But it brings up another question that needs further understanding and study that whether in multipartite EoP, there is any overcounting taking place that one needs to be careful about. Since we consider union of smaller black holes as our radiation state, a possible resolution in the purification side is simply considering that in the given limits (one black hole much much larger than all the other ones), the total state behaves as a bipartite state instead of a multipartite state and the entanglement of purification reduces to usual entanglement entropy.

\begin{equation}
  \Delta_{W} (A: B_{1}: B_{2}:....:B_{n})\longrightarrow  E_{P}(A:B) = S_{A} = S_{B},
\end{equation}
Where A is the bigger black hole and B is the union of the smaller black holes ($B_{1},B_{2},...,B_{n}$). To specify what is going to be $S_{A}=S_{B}$ at different times depends on which part of $\Delta_{W}$ is minimal choice. For example , we can divide $\Delta_{W}$ into two parts, one coming from the EWCS of two boundary wormhole ($\Delta_{W,1}=\sqrt{L_{0}^{2} - n \ell^{2}}$) and the other coming from the EWCS of the unions of the smaller boundaries ($\Delta_{W,2}=n \ell$). At all times, we can write,
\begin{equation}
    \Delta_{W} = \Delta_{W,1} + \Delta_{W,2},
\end{equation}

and at each time (for $n > 2$), in the limit where we consider the multipartite pure state as a bipartite one,
\begin{equation}
  E_{P}(A:B) = S_{A} = S_{B}= min(\Delta_{W,1}, \Delta_{W,2})\\\\
  =min(\sqrt{L_{0}^{2} - n \ell^{2}}, n \ell).
\end{equation}
This is a justifiable assumption since ultimately we are bothered about the entanglement between radiation state (union of smaller black holes) and the evaporating black hole state. Therefore it is not so unexpected that the initially multipartite situation reduces to simpler bipartite one. A detailed field-theoretic study similar to \cite{Balasubramanian:2014hda} in terms of purification would be able to shed more light on the necessity of this consideration. This is an ongoing problem that is in progress. Also, it would be interesting to consider different exits of a multiboundary case differently and study how would multipartite EoP behaves. 

5. At this point, it is also important to note that how a multipartite EoP is reduced to a bipartite case is very similar and pictorially same to choosing just the area of the HRT instead of choosing the area of the nontrivial island in this toy model. In both of the cases, one would encounter the information paradox had the alternate choice been made.

 While this was in preparation, \cite{Balapage} appeared online, which deals with multiboundary wormholes and purification from a different perspective. In the model introduced in \cite{Balapage}, they work with end of world branes and the multiboundary wormhole appears in the auxiliary system introduced for the purification. They consider something they term as ''inception geometry" to propose an extremal surface through which the nontrivial islands can again be marked. They argue that there are some region behind the horizon which can only be found if the Hawking radiation is considered as union of different subsystems of the radiation. They call such an event as quantum/geometric secret sharing. In our discussion through our resolution , we find that it is necessary to finally consider the system of big and smaller black holes as a bipartite pure state to make sense of the Page curve. But nevertheless, it is absolutely necessary to model numerous smaller black holes (subsystem Hawking radiation) to get the analog and intuitive understanding of islands. Had we just considered a bipartite pure state, i.e; a two boundary wormhole, we would never be able to get the shared interior which appears after the Page time. Note that in this case, there is only one choice in choosing the EWCS as well as the HRT. Therefore, our discussion in a way also addresses the necessity of modeling the radiation  as a union of subsystems as discussed in \cite{Balapage}. In light of such findings, we prefer to make the following statement,

\textit{Although multipartite purification in the multiboundary wormhole toy model gives back the overcounting once the boundary of the island is computed, it is absolutely necessary to primarily have the multipartite nature in the modeling of the radiation states to have a realisation of the islands in the toy model. To resolve the overcounting issue, we nevertheless need to review the model as the bipartite one and choose the minimal one among the two parts of the multipartite EWCS as the entanglement entropy of the bipartite pure state.}
 
Along with these lessons of how the two sides can feed each other with valuable new information, several future problems remain open that can be explored. The realization of both the multiboundary wormhole as well as multipartite entanglement of purification in higher space-time dimensions would be an interesting direction to pursue. A similar connection between multiparty EoP and MBW was made earlier in \cite{Bao:2018fso} but with a different goal. \footnote{We thank Aidan Chatwin-Davies for bringing this to our notice.} But the authors missed the subtlety of connecting subregions to boundaries when a handle is also involved in the MBW picture. It would be interesting to investigate whether such a description of islands in terms of multipartite EoP persists in higher dimensions and situations involving handles in the wormhole geometries. This would in principle mean that one will be able to make stronger statements about the connection between islands and quantum error correction. It is also important to note that since QEC relates entanglement wedge cross-sections to entanglement negativity, the connections between EoP and islands are also a topic worth studying. Since entanglement wedge cross-sections are also argued to pave the way towards a better understanding of full tensor networks in AdS/CFT, as advocated in \cite{Nguyen:2017yqw, Bao:2019fpq}, this can also be a potential candidate in providing us with a parallel understanding of the islands in terms of the tensor networks.

We believe this study will open up several directions that can be explored. These studies can potentially connect many of the recent exciting topics and strengthen our understanding in each of them.

\section*{Acknowledgements}
It is a pleasure to thank Shibaji Roy, Arnab Kundu, Harvendra Singh and other members of String Group in SINP for many useful discussions. We would also like to acknowledge Shibaji Roy, Arnab Kundu , Arpan Bhattacharyya and Avik Banerjee for useful comments and suggestions on the earlier draft of this work. Finally, I would like to thank the Department of Atomic Energy (DAE), Govt. of India for the funding.

\appendix
\section{More on Multiboundary Wormhole Construction:}\label{app A}
As mentioned in the main text, there are different methods of construction of multiboundary wormholes that have been discovered in different papers. In \cite{Caceres:2019giy}, a global three boundary wormhole construction has been discussed recently. The idea is to use the killing vectors of AdS$_{3}$ in the $t=0$ slice in Poincare coordinates. The exponentiated killing vectors are used as isometries in the identity component of SO$(2,2)$. The Killing vectors form an so$(2,2)$ algebra. In case of $t=0$ slice choice, the fixed points lie outside the fundamental domain and three of the six killing vectors are zero already.
\begin{figure}
	\[
	\begin{tikzpicture}[scale=0.7]
	\draw[->,very thin] (-5.5,0) to (5.5,0);
	\draw[->,very thin] (0,0) to (0,5.5);
	
	\draw[-,dashed,very thick,color=red] (0,1.25) to (0,5);
	\draw[-,dashed,very thick,color=blue] (41/12,0.372678) arc (41.81:138.19:0.559017);
	\draw[-,dashed,very thick,color=red!65!blue] (39/20,2/5) arc (36.87:118.07:2/3);
	\draw[-,dashed,very thick,color=red!65!blue] (4*75/68,4*10/17) arc (162.02:176.99:7.62209);
	
	\draw[-,color=red] (5,0) arc (0:180:5);
	\draw[-,color=red] (1.25,0) arc (0:180:1.25);
	
	\draw[-,color=blue] (1.75,0) arc (180:0:0.5);
	\draw[-,color=blue] (3.25,0) arc (180:0:0.5);
	
	\node[color=red,rotate=-90] at (0,1.25) {$\blacktriangle$};
	\node[color=red,rotate=-90] at (0,5) {$\blacktriangle$};
	
	\node[color=blue,rotate=-90] at (2.25,0.5) {$\blacktriangle$};
	\node[color=blue,rotate=90] at (3.75,0.5) {$\blacktriangle$};
	
	\draw (0,0) to (0,-0.2);
	\draw (2.25,0) to (2.25,-0.2);
	\draw (3.75,0) to (3.75,-0.2);
	
	\node at (0,-0.4) {$0$};
	\node at (2.25,-0.4) {$c_1$};
	\node at (3.75,-0.4) {$c_2$};
	
	\draw[->] (0,0) to (1.25/1.414,1.25/1.414);
	\node at (1.25/1.414-0.3,1.25/1.414+0.5) {$R_0$};
	
	\draw[->] (0,0) to (-5/1.414,5/1.414);
	\node at (-5/1.414+0.7,5/1.414-0.2) {$\lambda R_0$};
	
	\draw[->] (2.25,0) to (2.25-0.5/1.414,0.5/1.414);
	\node at (2.25,0.5/1.414+0.5) {$R$};
	
	\draw[->] (3.75,0) to (3.75+0.5/1.414,0.5/1.414);
	\node at (3.75,0.5/1.414+0.5) {$R$};
	
	\node at (0.3,3.125) {$L_1$};
	\node at (3,0.8) {$L_2$};
	\node at (1.5,1) {$L_{3}^L$};
	\node at (3.9,2) {$L_3^R$};
	\end{tikzpicture}
	\]
	\caption{Fundamental domain of the three-boundary Riemann surface $(3,0)$. The colored dashed lines $L_{1,2,3}$ are the minimal periodic geodesics, whose lengths are the physical parameters of the system. The variables $\lambda$, $R_0$, $R$, $c_1$, and $c_2$ represent parameters for the picture which are related to different killing vectors that transform the semicircles through isometries.}
	\label{figs:3bdrypic}
\end{figure}
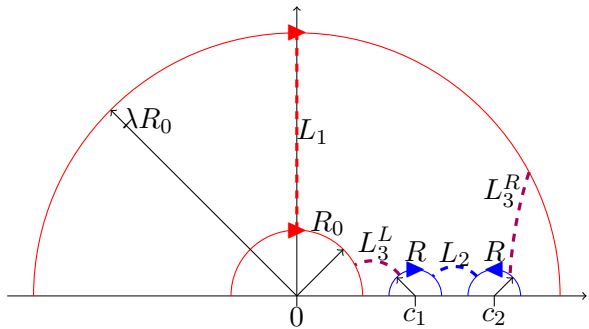

The Poincare metric of AdS$_{3}$ is written in the following way,
\begin{equation}
    ds^{2}= \frac{\ell^{2}}{p^{2}}\left( -dt^{2}+dx^{2}+dp^{2}\right)
\end{equation}
and the killing vectors in the $t=0$ slice are of the following form,

\begin{equation}
  J_{01}|_{t=0} = 0,
\end{equation}
\begin{equation}
  J_{02}|_{t=0} = 0,
\end{equation}
\begin{equation}
  J_{03}|_{t=0} = 0,
\end{equation}
\begin{equation}
  J_{12}|_{t=0} = -x \partial_{x} -p\partial_{p},
\end{equation}  
\begin{equation}
  J_{13}|_{t=0} = \left(\frac{\ell^{2}-x^{2}+p^{2}}{2\ell} \right)\partial_{x}-\frac{xp}{\ell}\partial_{p},
\end{equation}
\begin{equation}
  J_{23}|_{t=0} =\left(\frac{-\ell^{2}-x^{2}+p^{2}}{2\ell} \right)\partial_{x}-\frac{xp}{\ell}\partial_{p}.
\end{equation}
A basis change of $J_{13}$ and $J_{23}$ simplifies the action of these killing vectors or rather the isometries (the exponentiated version) on the complex upper half plane defined by ,
\begin{equation}
    z= x+ ip  \, , \, \bar{z}=x-ip.
\end{equation}
In the new basis $J_{12}$ acts as dilatation , $(J_{13}- J_{23})$ as translation and $(J_{13}-J_{23})$ as special conformal transformation. These and the inversion can provide all necessary transformations of a semicircle on the upper half plane to generate multiboundary wormhole with arbitrary number of boundaries(n) and genus(h). All connected, hyperbolic Riemann surfaces of genus h and boundaries n are denoted by (n,h) and for a particular value of n and h, the moduli space is known as the teichmuller space. It parametrically presents the number of geometrical parameters needed to construct a multiboundary wormhole. For a (2,0) wormhole, the number is just one (the horizon length of the 2 sided BTZ). Otherwise, it consists of $3h-3+2n$ number of minimal geodesics and $3h-3+n$ number of twist angles.

For a (3,0) wormhole, the number of parameters is thus just $3$ all of which correspond to the horizon lengths. These three horizon lengths $L_{1}$, $L_{2}$ $L_{3}$ are functions of mutually exclusive set of parameters (as found in \cite{Caceres:2019giy}) and hence their lengths can be tuned independently in the construction. 
\begin{figure}[t]
	\centering
	\begin{tikzpicture}[scale=0.7]
	\draw[->,thick] (-3.5,0) to (3.5,0);
	\draw[-,color=blue] (1,0) arc (180:0:0.75/2);
	\draw[-,color=blue] (2.75,0) arc (0:180:0.75/2);
	
	\draw[-,color=purple] (0.75,0) arc (0:180:0.75);
	\draw[-,color=purple] (3,0) arc (0:180:3);
	
	\node[color=blue,rotate=-90] at (1+0.75/2,0.75/2) {$\blacktriangle$};
	\node[color=blue,rotate=90] at (2.75-0.75/2,0.75/2) {$\blacktriangle$};
	
	\node[color=purple, rotate=-90] at (0,0.75) {$\blacktriangle$};
	\node[color=purple, rotate=-90] at (0,3) {$\blacktriangle$};
	
	\draw[->,very thick,dashed,purple] (0,3) to (0,0.75);
	\draw[->,very thick,dashed,blue] (1+0.75/2,0.5) to[bend left] (2.75-0.75/2,0.5);
	\draw[->,very thick,dashed,black] (0.3,0.55) to[bend left] (1+0.5/2,0.4);
	\draw[->,very thick,dashed,black] (2.75-0.45/2,0.35) to[bend left] (2.6,1.3);

	\draw[->,very thick] (4,1.75) to (8,1.75);
	
	\draw[-,color=black] (14.75,+1.75) arc (0:360:3);
	
	\draw[-,color=purple] (10,6-6.6867) to[bend left] (13.50,6-6.6867);
	\draw[-,color=purple] (10,6-1.8133) to[bend right] (13.50,6-1.8133);
	
	\draw[-,color=blue] (13.75,6-6.48607) to[bend left] (14.7,6-4.79544);
	\draw[-,color=blue] (13.75,6-2.01393) to[bend right] (14.7,6-3.70456);
	
	\draw[->,very thick,dashed,purple] (11.75,6-4.25+1.90) to (11.75,6-4.25-1.90);
	\draw[->,very thick,dashed,black] (12.50,-0.25) to[bend left] (13.9, 6-5.95);
	\draw[->,very thick,dashed,blue] (14.21, +0.65) to[bend left] (14.21, 6-3.35);
	\draw[->,very thick,dashed,black] (13.9, 6-2.65) to[bend left] (12.50,6-4.25+2.1);

	\end{tikzpicture}
	\caption{The three-boundary Riemann surface as quotients of the two-boundary Riemann surface. The left side of the picture shows the usual horizon choices in the upper half plane where the boundary of island consists of the dotted lines (horizons) and union of the solid lines between them. The right hand side of the picture shows the same region in the Poincare disk.}
	\label{figs:tikz2}
\end{figure}
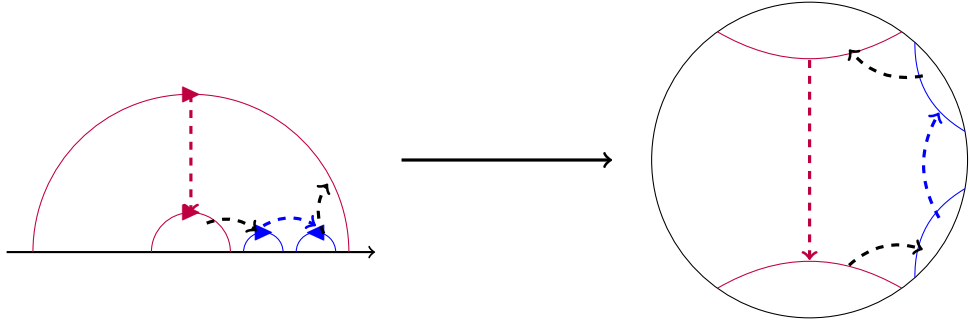
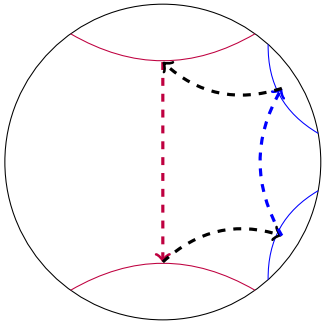
\begin{figure}[t]
	\centering
	\begin{tikzpicture}[scale=0.7]

	\draw[-,color=black] (14.75,-7+1.75) arc (0:360:3);

	\draw[-,color=purple] (10,-7.6867) to[bend left] (13.50,-7.6867);
	\draw[-,color=purple] (10,-2.8133) to[bend right] (13.50,-2.8133);
	
	\draw[-,color=blue] (13.75,-7.48607) to[bend left] (14.7,-5.79544);
	\draw[-,color=blue] (13.75,-3.01393) to[bend right] (14.7,-4.70456);
	
	\draw[->,very thick,dashed,purple] (11.75,-5.25+1.90) to (11.75,-5.25-1.90);
	\draw[->,very thick,dashed,black] (11.75,-5.25-1.90) to[bend left] (14.01, -6.65);
	\draw[->,very thick,dashed,blue] (14.01, -6.65) to[bend left] (14.01, -3.85);
	\draw[->,very thick,dashed,black] (14.01, -3.85) to[bend left] (11.75,-5.25+1.90);
	
	\end{tikzpicture}
	\caption{Multipartite EoP where the bulk minimal geodesics form a closed region. In case of large number of exits Figure \ref{figs:tikz2} matches with this kind of a closed bulk island.}
	\label{figs:tikz3}
\end{figure}
For our corresponding model to realise this from the subregion point of view, the same thing can be done by identifying the semicircles removed as the removed boundary subregions and the entanglement of purifications are then considered only for the regions in the remaining part of the boundary subregions which correspond to the actual boundaries in the multiboundary wormhole construction. The boundary anchored geodesics play the roles of the glueing surfaces (as discussed in Section 3 and Figures 6 and 7 of \cite{Bao:2018fso}).

The fundamental domain in the upper half plane would simply correspond to a decompactified Poincare disc picture for the subregions and the semicircles in Figure \ref{figs:3bdrypic} are the analogs of the boundary anchored geodesics again in a decompactified Poincare disc picture.

\section{Limitations and Clarifications:}\label{app B}
In this section, we list the precise limitations of our proposal.

1. The horizons in the multiboundary wormholes by themselves do not form a closed curve in the bulk in general as shown in Figure \ref{figs:tikz2}. It also includes part of the boundary anchored geodesics. But, in case of multipartite EoP, it is conjectured using the surface state correspondence that it is indeed a closed region (as shown in \ref{figs:tikz3}) in the bulk and doesn't involve any part from the boundary anchored geodesics except the points where different entanglement wedge cross sections meet each other. Hence the two pictures do not completely satisfy all the similarities in the most general situation. But, if we keep increasing the number of smaller exits, in the Poincare disk, we can easily see the part in the island that comes from intervals of the boundary anchored geodesics and not the horizons, keep decreasing. Thus in the large n limit (which is also the subject of interest in our model at later times), the parts of island coming from the boundary anchored geodesic intervals tend to zero and the two picture exactly correspond to each other in this limit.

2. In \cite{Bao:2018fso}, the authors compared EoP and MBW horizons to point out that they correspond to same things physically, but one is easier to compute in general than the other. But in our study we find that the multipartite EoP, which is the analogue of island, is not so easy to compute explicitly in general. But physically the island region also correspond to the region behind all of the horizons and thus, constitutes the entanglement shadow. Alhough it is hard to compute explicitly in general, which is a limitation, it provides reassuring physical hints that entanglement behind the horizon can be computed from the outside.

\bibliographystyle{JHEP}
\bibliography{single.bib}

\end{document}